\documentclass{aa}
\usepackage{psfig}

\begin{document}

\title{Unveiling the nature of {\it INTEGRAL} objects through optical
spectroscopy. III. Observations of seven southern 
sources\thanks{Based on observations collected at the South African 
Astronomical Observatory in Sutherland (South Africa) and at ESO, La 
Silla (Chile).}}

\author{N. Masetti\inst{1},
M.L. Pretorius\inst{2},
E. Palazzi\inst{1},
L. Bassani\inst{1},
A. Bazzano\inst{3}, 
A.J. Bird\inst{2},
P.A. Charles\inst{2,4}, \\
A.J. Dean\inst{2},
A. Malizia\inst{1},
P. Nkundabakura\inst{4,5},
J.B. Stephen\inst{1} and
P. Ubertini\inst{3}
}

\institute{
INAF -- Istituto di Astrofisica Spaziale e Fisica Cosmica di Bologna, 
Via Gobetti 101, I-40129 Bologna, Italy (formerly IASF/CNR,
Bologna)
\and
School of Physics \& Astronomy, University of Southampton, Southampton, 
Hampshire, SO17 1BJ, United Kingdom  
\and
INAF -- Istituto di Astrofisica Spaziale e Fisica Cosmica di
Roma, Via Fosso del Cavaliere 100, I-00133 Roma, Italy (formerly 
IASF/CNR, Roma)
\and
South African Astronomical Observatory, P.O. Box 9, Observatory 7935, 
South Africa 
\and
Department of Mathemathics \& Applied Mathemathics, University of Cape 
Town, Rondebosch 7701, South Africa
}

\offprints{N. Masetti (\texttt{masetti@iasfbo.inaf.it)}}
\date{Received 10 October 2005; accepted 15 December 2005}

\abstract{
In our continuing optical spectroscopic campaign to identify the
longer-wavelength counterparts of newly-discovered hard X--ray sources 
detected by {\it INTEGRAL}, we observed the putative 
optical counterparts of seven southern sources at the South African 
Astronomical Observatory and at the European Southern Observatory. 
For two of these objects, optical photometry was also acquired.
These observations firmly established the nature of four of them: 
we found that IGR J10404$-$4625 (=LEDA 93974), 4U 1344$-$60 and 
IGR J16482$-$3036 are Active Galactic Nuclei (AGNs)
at redshifts $z$ = 0.0237, 0.013 and 0.0313, respectively, 
and that 2RXP J130159.6$-$635806 is a Galactic High-Mass X--ray 
Binary (HMXB). We also give possible optical identifications for three 
further objects, namely IGR J11215$-$5952, IGR J11305$-$6256 and IGR 
J16207$-$5129, which are consistent with being Galactic HMXBs.
Physical parameters for these objects are also evaluated by collecting and 
discussing the available multiwavelength information. The detection of 
four definite or likely HMXBs out of seven objects in our sample 
further stresses {\it INTEGRAL}'s crucial contribution in hunting this 
class of object. Also, the determination of the extragalactic nature of a 
substantial fraction of the {\it INTEGRAL} survey sources 
underlines the importance of hard X--ray observations for the study of 
background AGNs located beyond the `Zone of Avoidance' of the Galactic 
Plane.

}

\titlerunning{Observations of 7 southern IGR sources}
\authorrunning{N. Masetti et al.}

\maketitle

\section{Introduction}

One of the objectives of the {\it INTEGRAL} mission (Winkler et al. 2003) 
is to survey the hard (20--100 keV) X--ray sky, with particular 
attention to the Galactic Plane. The capabilities of the IBIS 
instrument (Ubertini et al. 2003) allow {\it INTEGRAL} to detect hard 
X--ray sources at the mCrab level with a typical localization accuracy of 
2-3$'$ (Gros et al. 2003).
This combination of sensitivity and positional accuracy is unprecedented 
for surveys in the hard X--ray band above 20 keV and made it 
possible, for the first time, to resolve crowded regions of the hard 
X--ray sky such as the Galactic Centre and the spiral arms. These 
capabilities also allowed the discovery of many new hard X--ray 
extragalactic objects beyond the Galactic Plane (the so-called `Zone of 
Avoidance'), where interstellar obscuration hampers observations in soft 
X--rays.

Since the launch of {\it INTEGRAL} in October 2002, IBIS has already 
scanned large portions of the sky, and allowed the production of two deep 
(down to mCrab sensitivities) Galactic Plane Surveys (Bird et 
al. 2004, 2005; Bassani et al. 2004, 2005) as well as surveys of the 
Galactic Centre (Revnivtsev et al. 2004), of spiral arms of the Galaxy, 
such as the Sagittarius (Molkov et al. 2004) and Crux (Revnivtsev et al. 
2005) arms, and of the Coma cluster of galaxies (Krivonos et al. 2005).

Within these surveys, the ISGRI detector of IBIS revealed more than 200 
sources between 20 and 100 keV, the relative majority of them ($\sim$50\%) 
being Galactic X--ray binaries and with a smaller percentage ($\sim$10\%) 
of Active Galactic Nuclei (AGNs). However, about one fourth of the sample 
has no secure counterpart at longer wavelengths and therefore cannot yet 
be associated with any known class of objects. The 
majority of these unidentified sources are believed to be X--ray binaries, 
although some of them have been identified as AGNs (e.g., Masetti et al. 
2004, 2005; hereafter Papers I and II).

In order to reduce the {\it INTEGRAL} error circle, correlations 
with catalogues at longer wavelengths (soft X--ray, optical, near- and 
far-infrared, and/or radio) are employed. In particular, Stephen et al. 
(2005a,b) found a strong positional correlation between the ISGRI objects 
and the softer X--ray sources in the {\it ROSAT} catalogue of bright 
sources (Voges et al. 1999), showing that a bright {\it ROSAT} source, if 
present within an ISGRI error circle, is very likely the soft X--ray 
counterpart of the corresponding {\it INTEGRAL} object.
Similarly, Sazonov et al. (2005) accurately determined the positions 
of the soft X--ray counterparts of 6 {\it INTEGRAL} sources by using {\it 
Chandra} data. The use of the positional information coming from soft 
X--ray satellites therefore increases the positional accuracy to a few 
arcsecs, thus making the optical searches much easier.

Also, the presence of a radio object within the IBIS error box can
again be seen as an indication of an association between the radio
emitter and the {\it INTEGRAL} source (e.g., Paper I; Paper
II). However, whereas the cross-correlation with catalogues at other
wavebands is fundamental in pinpointing the putative optical
candidates, only optical spectroscopy can reveal the exact nature of
the X--ray emitting object.  Additionally, broadband optical
photometry can help to determine other characteristics, such as the
overall spectral energy distribution and absolute magnitude.

Thus, in our continuing effort to identify the unknown {\it INTEGRAL}
sources, here we concentrate on a sample of 7 southern objects for
which likely bright candidates could be pinpointed on the basis of
their association with sources in other wavebands.  These are mainly
in soft X--rays and radio, or for which a bright emission-line star
could be detected within the ISGRI error circle.

Admittedly, the latter cases are the weaker ones among our
selection of putative counterparts, due to the relatively large {\it
INTEGRAL} error boxes and the lack of more accurate catalogued
localizations at shorter wavelengths (other than optical). However, in
several instances a bright emission-line star in an {\it INTEGRAL}
error circle has been found to be the actual optical counterpart of
the source responsible for the detected hard X--ray emission (e.g.,
Reig et al. 2005), so we consider these stars as important and viable
candidates. We refer the reader to Reig et al. (2005) for a detailed
discussion of the possibilities and caveats that these putative
associations imply.

We moreover remark the following: using the spectral information 
of the stars in the Hipparcos catalogue (Perryman et al. 1997), together 
with (i) the known number densities of bright stars, (ii) the proportion 
of early-type stars showing emission lines and (iii) the percentage of 
stars in each spectral class (as in Allen 1973), we find that, in our 
cases, the chance probability of finding a bright blue emission-line 
star within the {\it INTEGRAL} error box is less than 0.05 \% along the 
Galactic Plane. This makes us confident that our choice of putative 
optical bright candidates of early spectral type and with emission lines
is fully justified.

We present here the spectroscopic results on these 7 sources obtained
at the South African Astrophysical Observatory (SAAO) and at the
European Southern Observatory (ESO). For two of them, optical
photometry was also obtained, and is reported here.
In Sect. 2 we present the sample of objects selected for this
observational campaign, whereas in Sect. 3 a description of the
observations is given; Sect. 4 reports the results for each source and
discusses their nature. Conclusions are drawn in Sect. 5.

In the following, when not explicitly stated otherwise, for our X--ray
flux estimates we will assume a Crab-like spectrum.

\section{The selected sample}

{\it IGR J10404$-$4625}: this source is present in the 2$^{\rm nd}$
IBIS survey (Bassani et al. 2005; Bird et al. 2005) and is detected
with fluxes of 2.9$\pm$0.7 mCrab and 5.9$\pm$1.1 mCrab in the 20--40
and 40--100 keV bands, respectively. Inside the {\it INTEGRAL} error
box no bright soft X--ray sources are found; however, a single radio
source, SUMSS\footnote{the SUMSS catalogue is available at \\ {\tt
http://www.astrop.physics.usyd.edu.au/SUMSS/}} J104022$-$462525, with
flux density 61.3$\pm$2.0 mJy at 843 MHz, is present.  This is in turn
positionally coincident with a far-infrared {\it IRAS} source and with
the optical S0-a type galaxy LEDA 93974 (Paturel et al. 2003; Fig. 1,
upper left panel). This latter object is classified as an
emission-line spiral galaxy at redshift $z$ = 0.024 (Wamsteker et
al. 1985; Stein 1996), but no optical spectral classification is
available at present, nor has any spectrum been published.
Nevertheless, the reported results suggest that LEDA 93974 is an
active galaxy.

{\it IGR J11215$-$5952}: this variable, possibly transient, source was 
detected in the 20--60 keV band on 2005 April 22, during an 
{\it INTEGRAL} Galactic Plane scan, at a flux of 75 mCrab 
which halved $\sim$40 minutes later (Lubinski et al. 2005). The source 
was not detected in the 60--200 keV band; this, according to those 
authors suggests a soft spectrum for the object. As noted by 
Negueruela et al. (2005), the B1\,Ia-type supergiant HD 306414 
(Vijapurkar \& Drilling 1993) lies well inside the {\it INTEGRAL} 
error circle of IGR J11215$-$5952 (Fig. 1, upper middle panel). On the 
basis of the photometric properties of this source, Negueruela et al. 
(2005) suggested this star as the optical counterpart of the 
{\it INTEGRAL} hard X--ray source and that its distance is around 8 kpc. 

{\it IGR J11305$-$6256}: this previously unknown hard X--ray transient
(Produit et al. 2004) was discovered by IBIS/ISGRI on May 12, 2004
during an {\it INTEGRAL} observation of the Carina region.  Produit et
al. (2004) report a detection at an average flux of 8 mCrab in the 20--60 
keV band, while there was no detection in the 3--10 keV band (no
value for the upper limit to the X--ray flux in this band is however
available). The only conspicuous catalogued object within the {\it
INTEGRAL} error box of IGR J11305$-$6256 is the emission-line star HD
100199 (Fig. 1, upper right panel), which has $V$ = 8.23 and $B-V$ =
+0.01 (Fernie 1983).  This is classified as a blue giant of spectral
type B0\,IIIe (Garrison et al. 1977), but no spectrum has been
published, and no further information is available on this
star. However, its identification as an early-type emission-line star
suggests HD 100199 as a strong candidate for IGR J11305$-$6256, by
analogy with other High-Mass X--ray Binaries (HMXBs) detected with
{\it INTEGRAL} (e.g., Reig et al. 2005; Paper II), even if the
relatively large error circle (5$'$) makes this identification less
secure.

{\it 2RXP J130159.6$-$635806}: known also as IGR J13020$-$6359, this
source was detected in both the Crux arm survey (Revnivtsev et al.
2005) and in the 2$^{\rm nd}$ IBIS survey (Bird et al. 2005). In the 
latter survey it appears with fluxes of 2.1$\pm$0.2 mCrab and 1.3$\pm$0.4
mCrab in the 20-40 keV and 40--100 keV bands, respectively.  At the
edge of the ISGRI error circle, a {\it ROSAT} and {\it XMM-Newton}
source was found (Chernyakova et al. 2005) to be an accreting X--ray
pulsar with spin period $\sim$700 s and spectral characteristics of a
HMXB. The smaller {\it XMM-Newton} error box (3$''$ radius;
Chernyakova et al. 2005) encompasses two optical objects (Fig. 1,
central left panel), a brighter one with USNO-A2.0\footnote{available
at \\ {\tt http://archive.eso.org/skycat/servers/usnoa/}} magnitude $R
\sim$ 13.9 and a fainter one of unknown magnitude; for the latter 
object, a rough magnitude estimate using the relevant DSS-II-Red Optical
Survey\footnote{available at \texttt{http://archive.eso.org/dss/dss/}} 
image gives $R \approx$ 17. Optical spectroscopy is therefore needed to 
clarify which of the two is the actual counterpart to 2RXP 
J130159.6$-$635806, and to independently confirm the HMXB nature of this 
source.

{\it 4U 1344$-$60}: this Ariel and Uhuru source (Seward et al. 1976;
Forman et al. 1978) was detected, for the first time above 10 keV, in
the 2$^{\rm nd}$ IBIS survey (Bird et al. 2005), at 20--40 and 40--100 keV
fluxes of 3.9$\pm$0.2 mCrab and 4.7$\pm$0.4 mCrab, respectively. It is
also associated with a source seen by {\it Einstein}, {\it EXOSAT},
{\it ASCA} and {\it XMM-Newton}. The data from {\it Einstein}
(McDowell 1994) and {\it EXOSAT} (Warwick et al. 1988) indicate that
this source has fluxes of $\sim$2$\times$10$^{-12}$ erg cm$^{-2}$
s$^{-1}$ and $\sim$2.3$\times$10$^{-11}$ erg cm$^{-2}$ s$^{-1}$ in the
0.16--3.5 keV and 2--6 keV bands, respectively. According to Michel et
al. (2004) and Bassani et al.  (2005), this source has the X--ray
spectral characteristics of an AGN. But, despite this long history of
observations, the object is still optically unidentified. However, the
smaller {\it XMM-Newton} error box (4$''$ radius) with coordinates RA =
13$^{\rm h}$ 47$^{\rm m}$ 36$\fs$43, Dec = $-$60$^{\circ}$ 37$'$
02$\farcs$4 (J2000) available in a serendipitous archival observation
of this field\footnote{available at \\ {\tt
http://heasarc.gsfc.nasa.gov/docs/archive.html}} allowed us to
pinpoint two optical objects consistent with the {\it XMM-Newton}
position (see Fig. 1, central middle panel), the brighter one at $R
\sim$ 16.4 according to the USNO-A2.0 catalogue, and the fainter one
of unknown magnitude; in this case also, we roughly estimate its 
magnitude as $R \approx$ 19 from the corresponding DSS-II-Red Survey 
image. As for IGR J13020-6359, optical spectroscopy of both objects is 
now required to complete this identification.

{\it IGR J16207$-$5129}: this hard X--ray source was detected for the
first time by IBIS/ISGRI in the 1$^{\rm st}$ Galactic Plane Survey
performed with {\it INTEGRAL} (Bird et al. 2004), with a flux of
3.8$\pm$0.3 mCrab in the 20--40 keV band; only an upper limit of
$<$4 mCrab was instead obtained in the 40--100 keV band. Again, an
emission-line star, HD 146803, is the only remarkable catalogued
object present inside the {\it INTEGRAL} error box (Fig. 1, central
right panel). This star shows H$_\alpha$ in emission (MacConnell 1981) 
and is of spectral type A1\,IV (Houk 1978). According to the 
information catalogued in SIMBAD\footnote{see
{\tt http://simbad.u-strasbg.fr/}}, HD 146803 has $B$ = 10.41 and $V$
= 10.45. The spectral characteristics of this star again, as for the
cases above, strongly suggest it to be the optical counterpart
of the hard X--ray source detected by {\it INTEGRAL}.

{\it IGR J16482$-$3036}: this object also has been detected in the
2$^{\rm nd}$ IBIS survey (Bassani et al. 2005; Bird et al.
2005), with fluxes of 1.6$\pm$0.2 mCrab (20--40 keV) and 1.6$\pm$0.3
mCrab (40--100 keV). Inside the {\it INTEGRAL} error box the soft
X--ray source 1RXS J164815.5$-$303511, belonging to the {\it ROSAT}
Bright Source Catalogue (Voges et al. 1999), is found at a 0.1--2.4
keV flux of (8.3$\pm$1.3)$\times$10$^{-13}$ erg cm$^{-2}$
s$^{-1}$. This, according to Stephen et al. (2005a,b), strongly
suggests that this source is the soft X--ray counterpart of IGR
J16482$-$3036. Consistent with the {\it ROSAT} position, a NVSS radio
source, with flux of 3.5$\pm$0.6 mJy at 1.4 GHz (Condon et al. 1998),
is present. The intersection of the smaller NVSS and {\it ROSAT} error
boxes allows us to pinpoint an extended optical object on the
DSS-II-Red Survey image (Fig. 1, lower left panel), which has $R 
\sim$ 13.4 and $B \sim$ 15.8 according to the USNO-B1.0 catalogue (Monet 
et al. 2003). Optical spectroscopy would thus conclusively determine the 
nature of this source.

\begin{figure*}
\hspace{-.5cm}
\centering{\mbox{\psfig{file=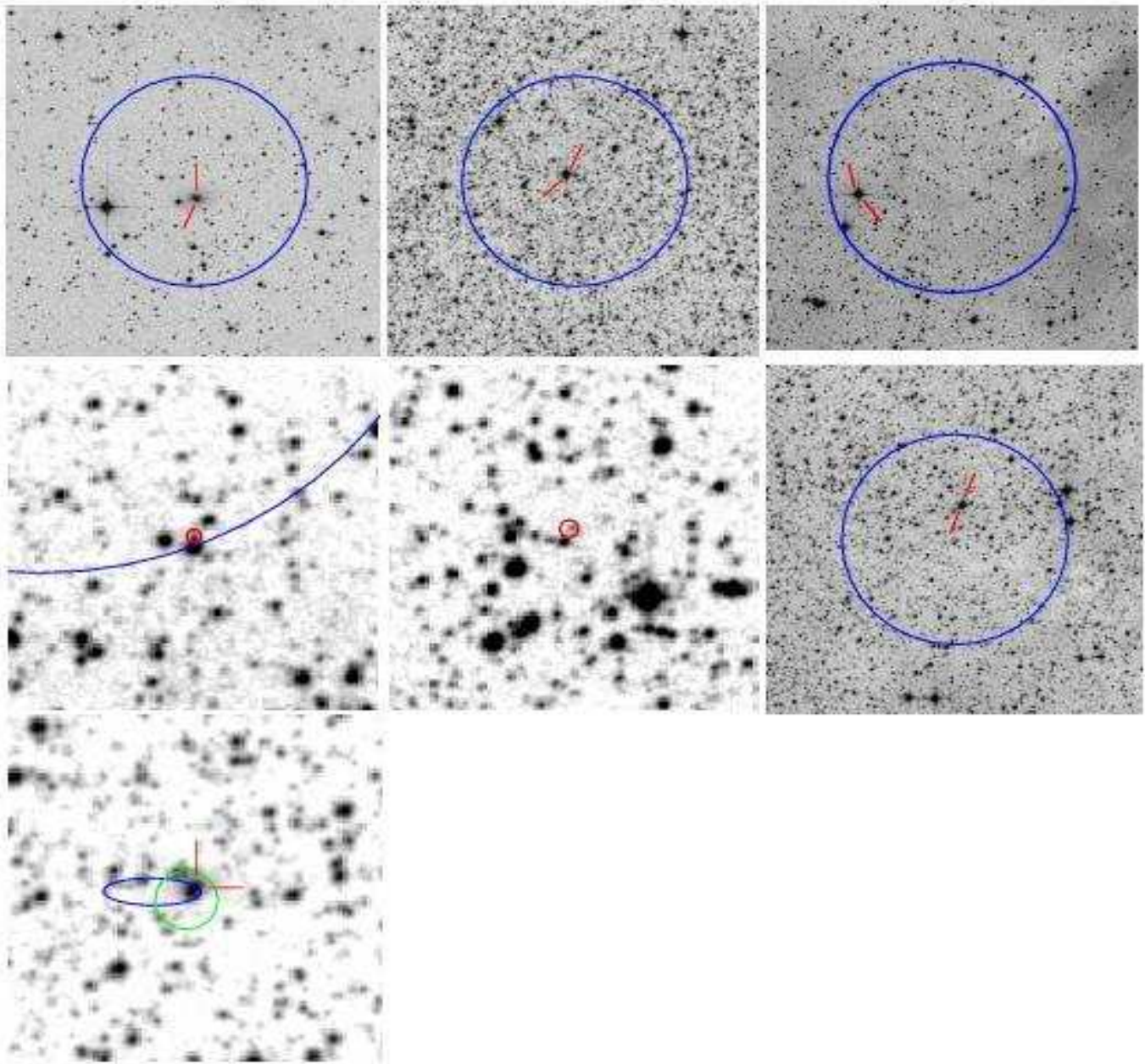,width=18cm}}}
\caption{DSS-II-Red optical images of the fields of IGR J10404$-$4625 
(upper left panel), IGR J11215$-$5952 (upper middle panel), IGR 
J11305$-$6256 (upper right panel), 2RXP J130159.6$-$635806 (centre 
left panel), 4U 1344$-$60 (central middle panel), IGR J16207$-$5129 
(central right panel) and IGR J16482$-$3036 (lower left panel).
For the objects in the upper panels, and for IGR J16207$-$5129, circles 
mark the ISGRI/{\it INTEGRAL} error boxes of the hard X--ray sources.
The putative optical counterparts are indicated with tick marks or,
for 2RXP J130159.6$-$635806 and 4U 1344$-$60, by the small 
{\it XMM-Newton} error circles (in the case of 2RXP 
J130159.6$-$635806 only a portion of the larger {\it INTEGRAL} error box 
is shown). For IGR J16482$-$3036 the light circle indicates the 
{\it ROSAT} soft X--ray position, whereas the ellipse marks the NVSS 
radio error box.
Field sizes are 15$'$$\times$15$'$ for IGR J11305$-$6256,
10$'$$\times$10$'$ for IGR J10404$-$4625, IGR J11215$-$5952 
and IGR J16207$-$5129, and 2$\farcm$5$\times$2$\farcm$5 for 2RXP 
J130159.6$-$635806, 4U 1344$-$60 and IGR J16482$-$3036.
In all cases, North is up and East to the left.}
\end{figure*}

\section{Observations at SAAO and at ESO}

\subsection{Spectroscopy}

\begin{table*}[t!]
\caption[]{Log of the observations presented in this paper.}
\begin{center}
\begin{tabular}{llccccc}
\noalign{\smallskip}
\hline
\noalign{\smallskip}
\multicolumn{1}{c}{Object} & \multicolumn{1}{c}{Date} & Mid-exposure & 
Telescope & Grism & Slit or & Exposure \\
 & & time (UT) & & or filter & seeing ($''$) & time (s) \\
\noalign{\smallskip}
\hline
\noalign{\smallskip}
\multicolumn{7}{c}{Spectroscopy} \\
\noalign{\smallskip}
\hline
\noalign{\smallskip}

 LEDA 93974 & 23 Jul 2005 & 17:25:24 & SAAO 1.9m & \#7 & 1.8 & 1000 \\

 2RXP J130159.6$-$635806 & 28 Jun 2003 & 02:34:04 & ESO 3.6m & \#6 & 1.5 & 
2$\times$600 \\

 4U 1344$-$60 & 01 Mar 2003 & 08:09:36 & ESO 3.6m & \#6 & 1.5 & 1000 \\

 HD100199 & 23 Jul 2005 & 17:58:28 & SAAO 1.9m & \#7 & 1.8 & 90 \\

 HD306414 & 23 Jul 2005 & 17:45:05 & SAAO 1.9m & \#7 & 1.8 & 300 \\

 HD146803 & 22 Jul 2005 & 19:22:10 & SAAO 1.9m & \#7 & 1.8 & 200 \\

 IGR J16482$-$3036 & 22 Jul 2005 & 19:44:10 & SAAO 1.9m & \#7 & 1.8 & 1000 \\

\noalign{\smallskip}
\hline
\noalign{\smallskip}
\multicolumn{7}{c}{Imaging} \\
\noalign{\smallskip}
\hline
\noalign{\smallskip}

 2RXP J130159.6$-$635806 & 28 Jun 2003 & 02:19:26 & ESO 3.6m & R & 1.6 & 20 \\

 2RXP J130159.6$-$635806 & 28 Jun 2003 & 02:47:45 & ESO 3.6m & i & 1.9 & 300 \\

 2RXP J130159.6$-$635806 & 28 Jun 2003 & 02:53:27 & ESO 3.6m & V & 2.1 & 300 \\

 4U 1344$-$60 & 01 Mar 2003 & 07:51:40 & ESO 3.6m & R & 1.0 & 20 \\

\noalign{\smallskip}
\hline
\noalign{\smallskip}
\end{tabular}
\end{center}
\end{table*}

We observed five southern optical objects of our sample with the SAAO 
1.9-metre ``Radcliffe" telescope located near Sutherland, South Africa. 
The Radcliffe telescope carried a CCD spectrograph mounted at the 
Cassegrain focus; the instrument was equipped with a 1798$\times$266 
pixel SITe CCD. In all observations, Grating \#7 and a slit of 
1$\farcs$8 were used, providing a 3850--7200 \AA~nominal spectral 
coverage. This setup gave a dispersion of 2.3~\AA/pix for all spectra 
acquired at SAAO.

We also retrieved from the ESO archive\footnote{available at {\tt
http://archive.eso.org/}} spectroscopic observations of the optical
candidates of both 4U1344$-$60 (ESO Program ID: 70.D-0227) and 2RXP
J130159.6$-$635806 (ESO Program ID: 71.D-0296). These were obtained
with the 3.6m ESO telescope located in La Silla (Chile); this
telescope was equipped with EFOSC2 and a 2048$\times$2048 pixel
Loral/Lesser CCD with 2$\times$2 binning.  For both pointings Grism
\#6 and a slit of 1$\farcs$5 were used, giving a 3850--8100
\AA~nominal spectral coverage, at a dispersion of 4.1~\AA/pix. In
these two cases the slit was oriented in such a way as to include both
candidates.  The complete spectroscopy observation log for both SAAO
and ESO is contained in Table 1.

The spectra, after cosmic-ray rejection and correction for bias and 
flat field, were reduced, background subtracted and optimally 
extracted (Horne 1986) using IRAF\footnote{IRAF is the Image Analysis 
and Reduction Facility made available to the astronomical community by 
the National Optical Astronomy Observatories, which are operated by 
AURA, Inc., under contract with the U.S. National Science Foundation. 
It is available at {\tt http://iraf.noao.edu/}}. Wavelength 
calibration was performed using Cu-Ar lamps for the SAAO spectra and 
He-Ar lamps for the ESO spectra. SAAO spectra were then 
flux-calibrated by using the spectrophotometric standards 
CD-32$^\circ$9927 and LTT 377, whereas those secured at ESO were 
calibrated in flux using the standard LTT9239; these three standards are 
extracted from the catalogues of Hamuy et al. (1992, 1994). In the 
case of 2RXP J130159.6$-$635806, we stacked together the two spectra 
to increase the signal-to-noise ratio (S/N). The wavelength calibration 
uncertainty was $\sim$0.5~\AA~for all cases, and was checked by using 
the positions of background night sky lines.

\subsection{Photometry}

Through the ESO archive we also retrieved optical imaging frames of 
the fields of 2RXP J130159.6$-$635806 and of 4U 1344$-$60, acquired 
again with EFOSC2 at the ESO 3.6m under the same programs and on the same 
nights in which spectroscopy was performed. In detail, a single $R$-band 
frame was available for the 4U 1344$-$60 field, and $VRi$ images for the 
field of 2RXP J130159.6$-$635806. Table 1 also contains information on 
these imaging observations.

The 2$\times$2-rebinned CCD of EFOSC2 secured a plate scale of
0$\farcs$31/pix, and a useful field of 5$\farcm$2$\times$5$\farcm$2. 
Images were corrected for bias and flat-field in the usual fashion and 
calibrated using the EFOSC2 zero-points\footnote{available at: \\ {\tt
http://www.ls.eso.org/lasilla/sciops/3p6/efosc/zp/}}.  The putative
counterparts of these two {\it INTEGRAL} sources were clearly detected
in all images, and their magnitudes were determined with
MIDAS\footnote{\texttt{http://www.eso.org/projects/esomidas}}.
Because of (i) the pointlike appearance of all candidates and (ii) the
fairly crowded fields, we applied the {\sl DAOPHOT} PSF-fitting
routine (Stetson 1987).

\begin{figure*}[th!]
\centering{\mbox{\psfig{file=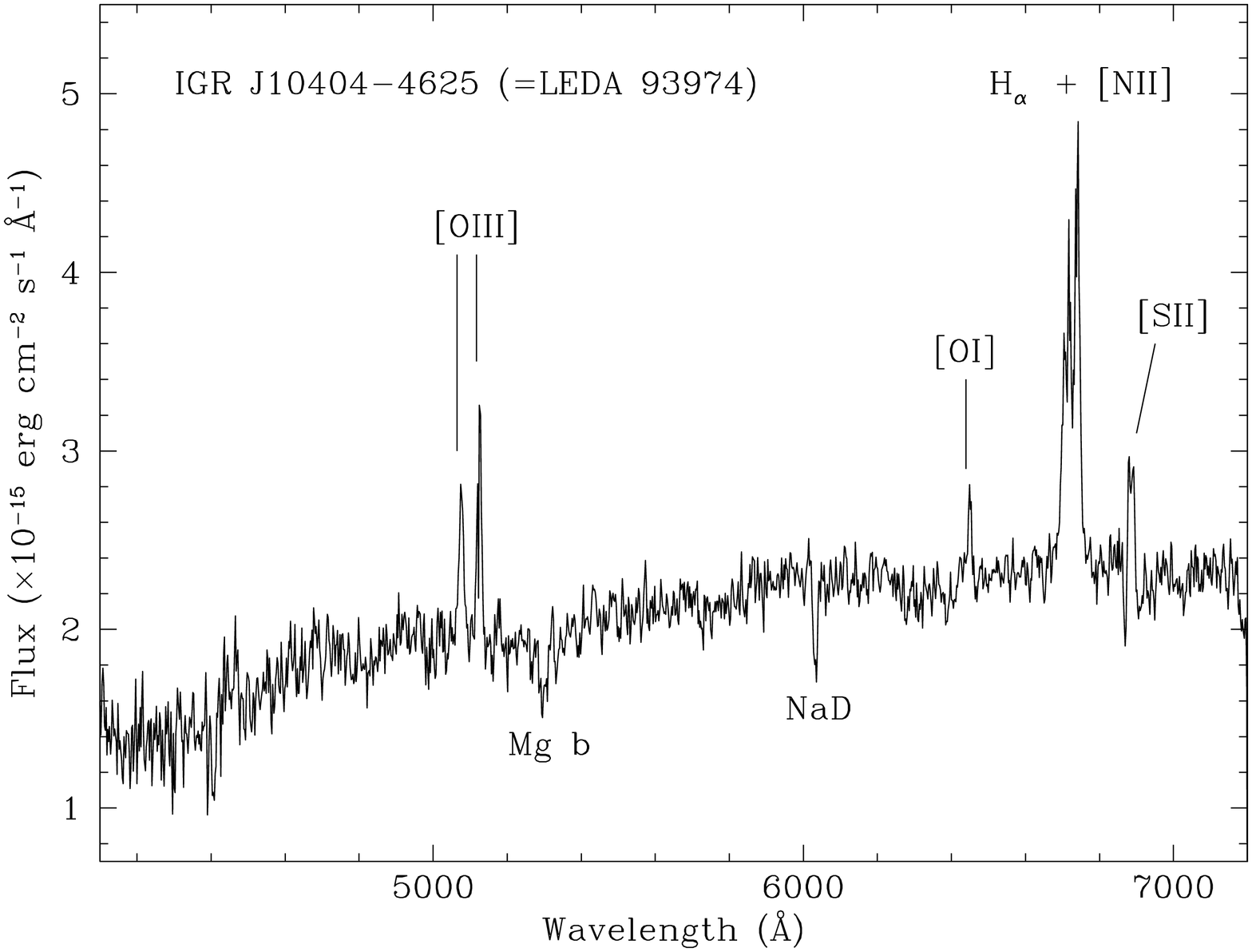,width=8cm,angle=270}}}
\centering{\mbox{\psfig{file=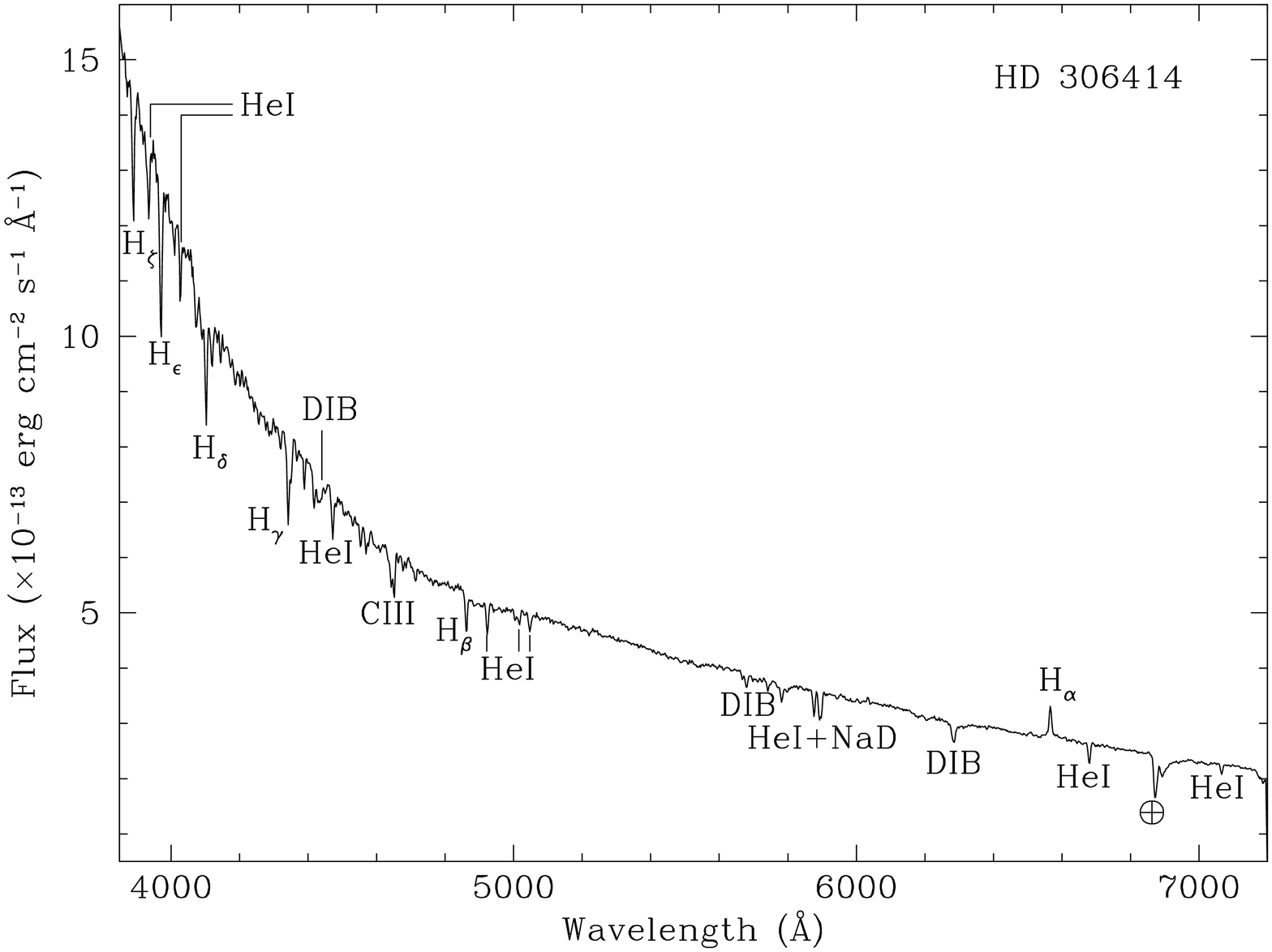,width=8cm,angle=270}}}
\centering{\mbox{\psfig{file=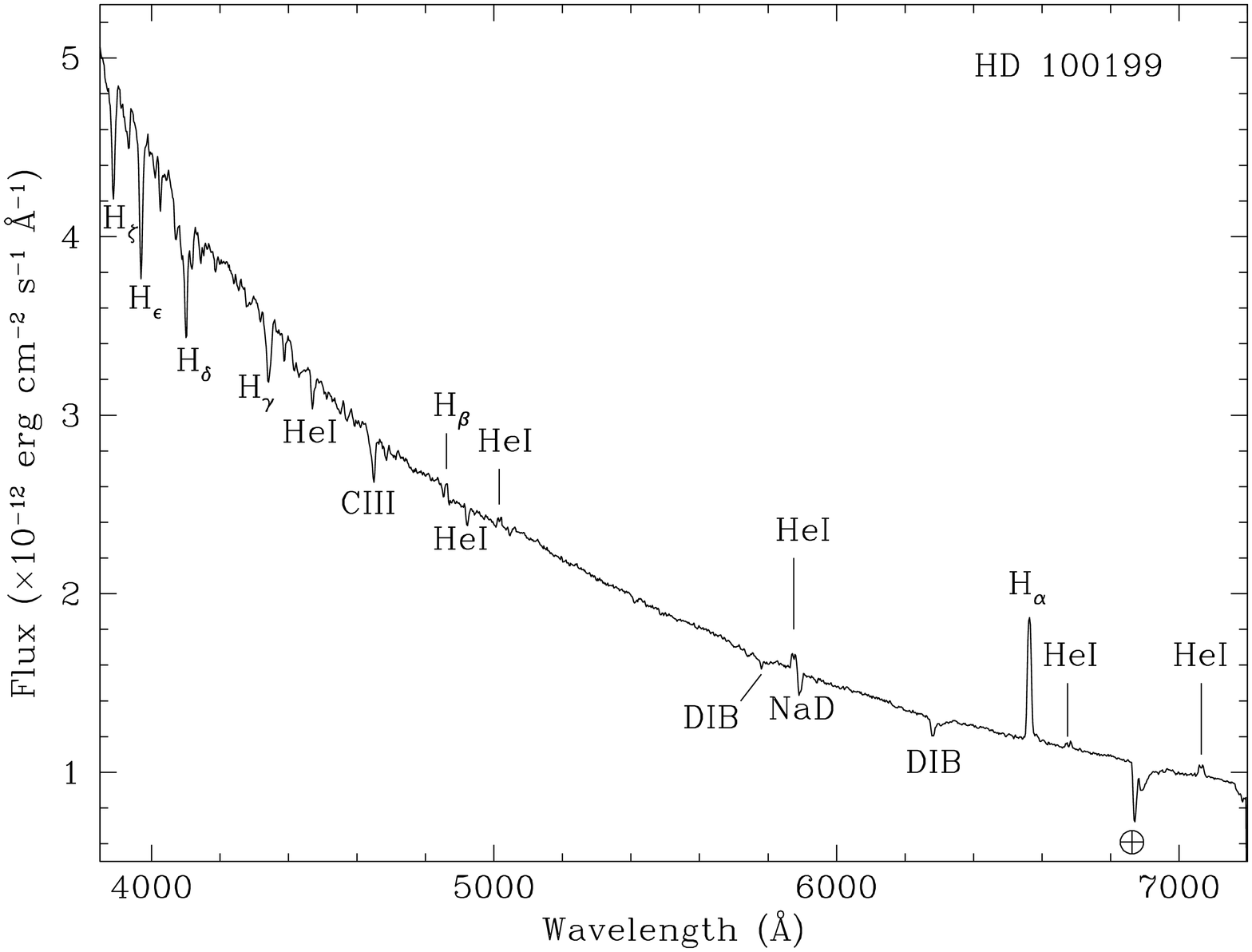,width=8cm,angle=270}}}
\centering{\mbox{\psfig{file=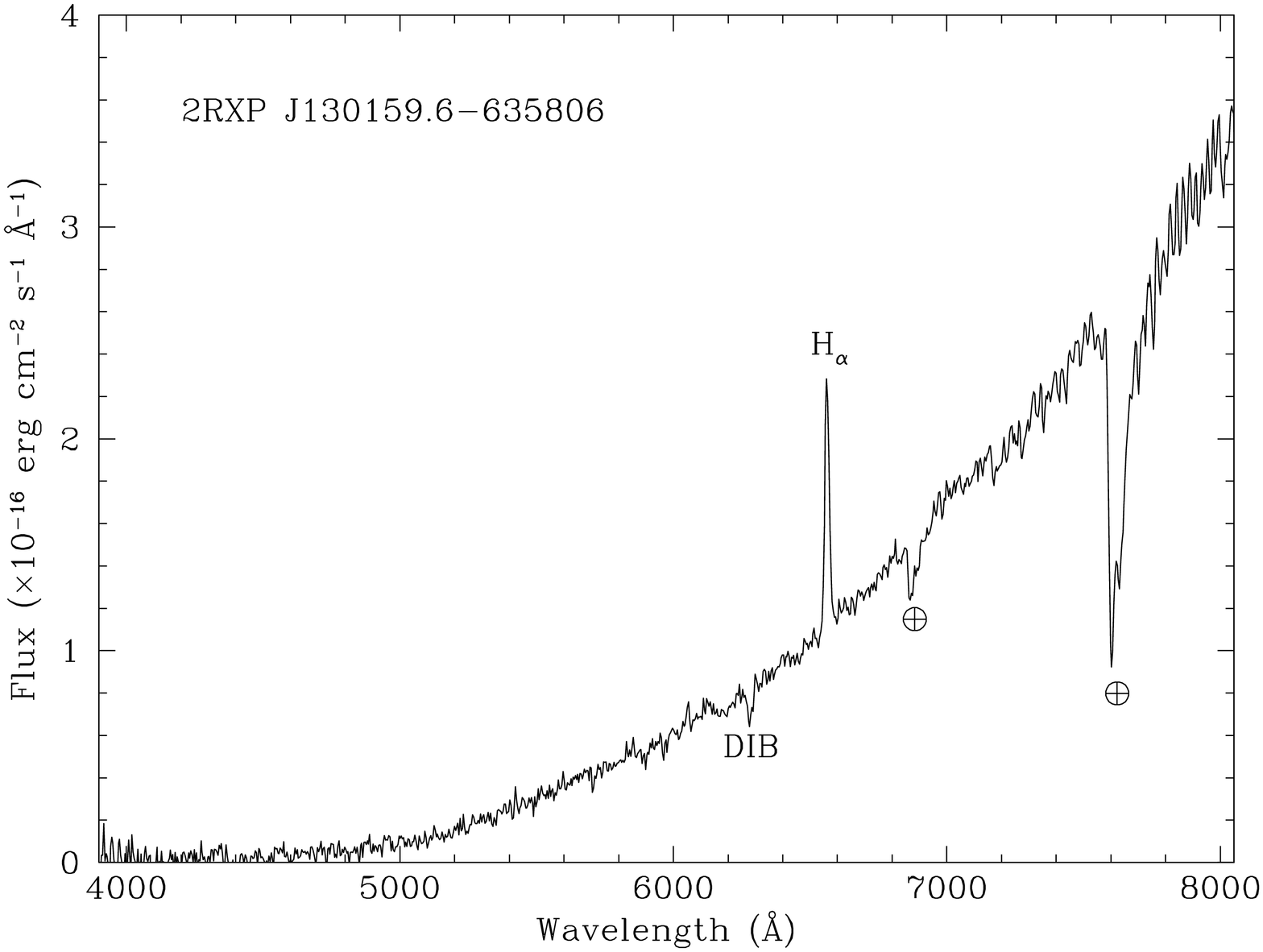,width=8cm,angle=270}}}
\centering{\mbox{\psfig{file=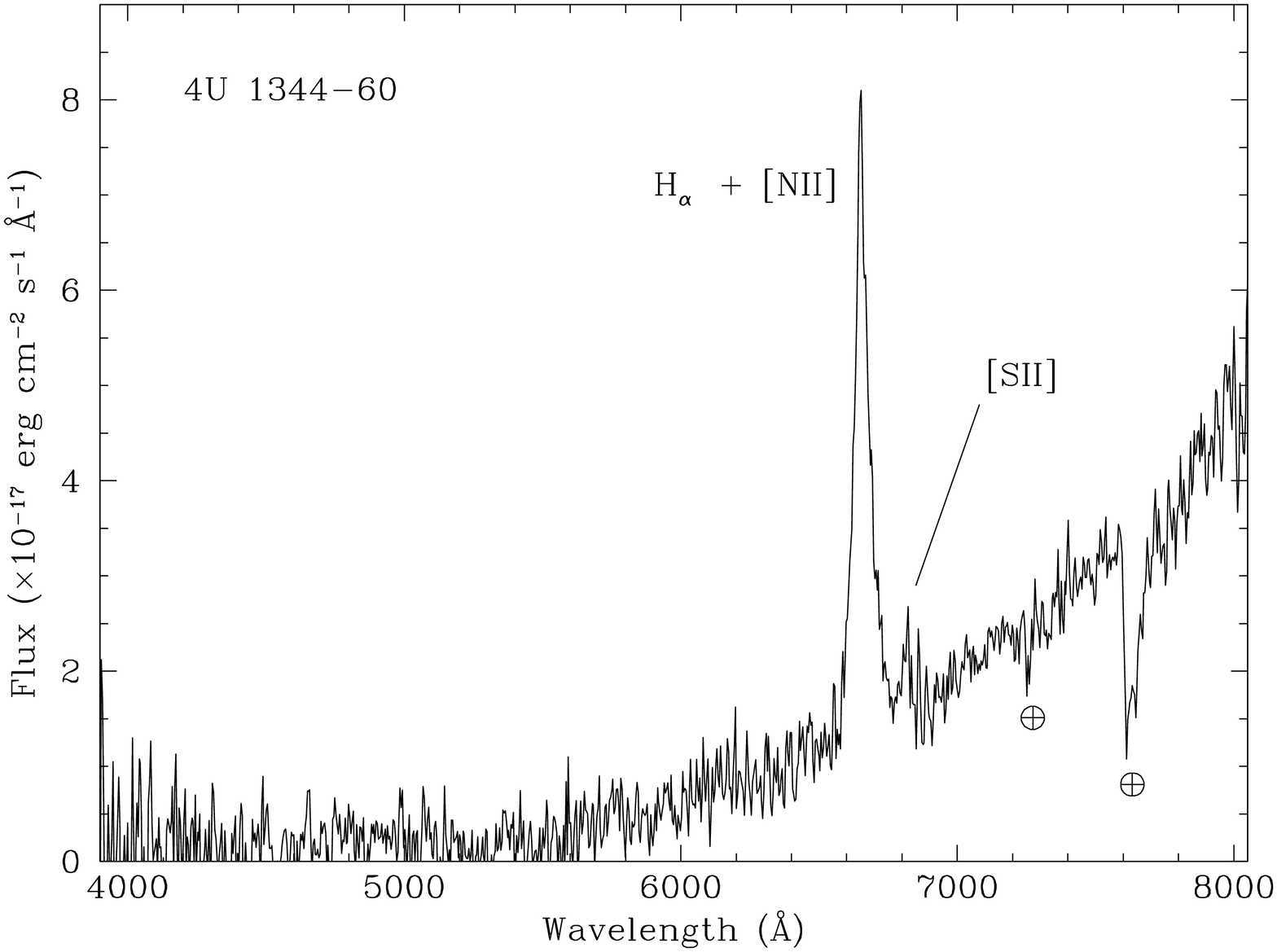,width=8cm,angle=270}}}
\centering{\mbox{\psfig{file=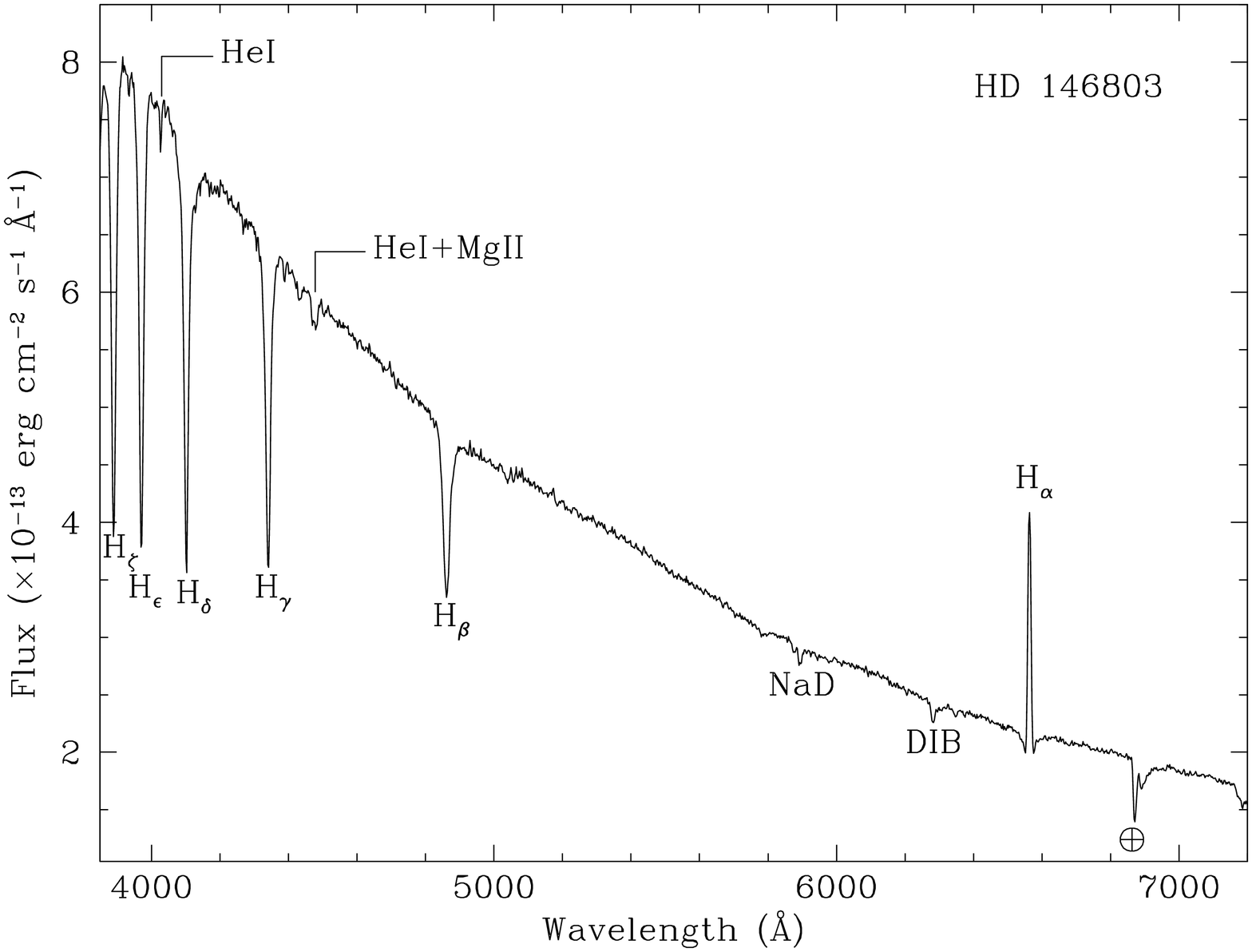,width=8cm,angle=270}}}
\begin{center}
\parbox{8cm}{
\hspace{-.2cm}
\psfig{file=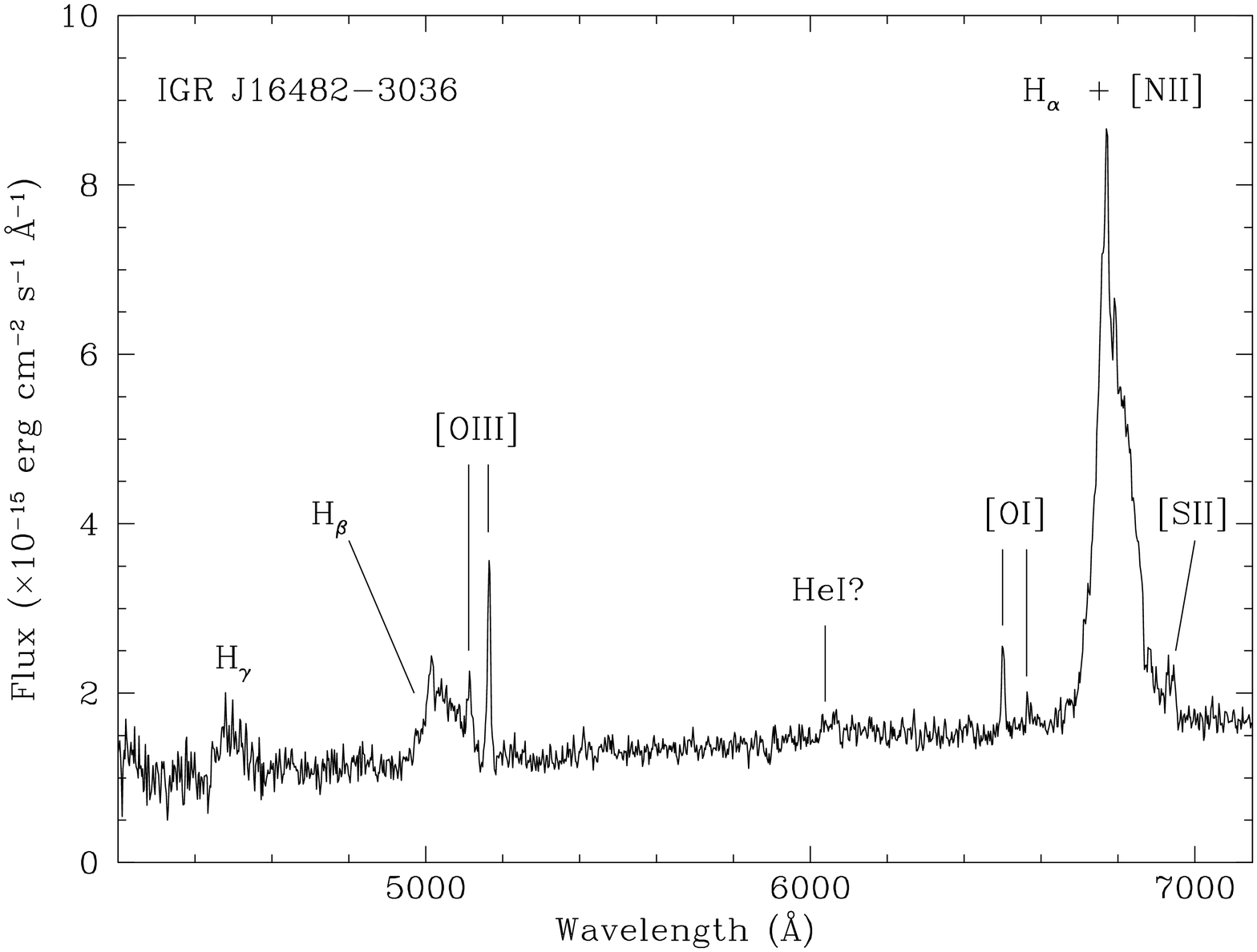,width=8cm,angle=270}
}
\parbox{7.8cm}{
\vspace{-1cm}
\caption{Optical spectra (not corrected for interstellar extinction) of 
IGR J10404$-$4625 (=LEDA 93974), HD 306414, HD 100199, 2RXP 
J130159.6$-$635806, 4U 1344-60, HD146803 and IGR J16482$-$3036. For each 
spectrum the main features are labeled. The symbol $\oplus$ indicates 
atmospheric telluric bands.}
}
\end{center}
\end{figure*}

\section{Results and discussion}

Table 2 contains the (observer's frame) emission-line wavelengths,
fluxes and equivalent widths (EWs) for the seven spectra displayed in
Fig. 2.  The line fluxes from extragalactic objects were dereddened
for Galactic absorption along the respective lines of sight following
the prescription of Schlegel et al. (1998) and assuming the Galactic
extinction law of Cardelli et al. (1989). These same spectra were not
corrected for starlight contamination (see, e.g., Ho et al. 1993,
1997) given the limited S/N and resolution of the spectrum. We do not
expect that this affects any of our conclusions.  In the following we
assume a cosmology with $H_{\rm 0}$ = 65 km s$^{-1}$ Mpc$^{-1}$,
$\Omega_{\Lambda}$ = 0.7 and $\Omega_{\rm m}$ = 0.3.

\begin{table}[th!]
\caption[]{Observer's frame wavelengths, EWs and fluxes (in units of
10$^{-15}$ erg s$^{-1}$ cm$^{-2}$) of the emission lines detected in the 
spectra of the seven objects shown in Fig. 2.
For the extragalactic sources the values are corrected for Galactic
extinction assuming, from Schlegel et al. (1998), $E(B-V)$ = 0.16 (for
LEDA 93974), 2.90 (for 4U 1344$-$60) and 0.34 (for IGR
J16482$-$3036). The error on the line positions is conservatively
assumed to be $\pm$2 \AA~and $\pm$4 \AA~for the SAAO and ESO spectra,
respectively, i.e. comparable to the corresponding spectral
dispersions (see text).} 
\begin{center}
\begin{tabular}{lcrr}
\noalign{\smallskip}
\hline
\noalign{\smallskip}
\multicolumn{1}{c}{Line} & $\lambda_{\rm obs}$ (\AA) & 
\multicolumn{1}{c}{EW$_{\rm obs}$ (\AA)} & \multicolumn{1}{c}{Flux} \\
\noalign{\smallskip}
\hline
\noalign{\smallskip}

\multicolumn{4}{c}{IGR J10404$-$4625 (= LEDA 93974)} \\
                                                                      
$[$O {\sc iii}$]$ $\lambda$4958 & 5076 &  7.3$\pm$0.7 &   23$\pm$2 \\
$[$O {\sc iii}$]$ $\lambda$5007 & 5125 &  8.6$\pm$0.9 &   27$\pm$3 \\
$[$O {\sc i}$]$ $\lambda$6300   & 6450 &  2.4$\pm$0.4 &  7.9$\pm$1.2 \\
$[$N {\sc ii}$]$ $\lambda$6548  & 6705 &  8.6$\pm$0.9 &   29$\pm$3  \\
H$_\alpha$                      & 6720 &  7.1$\pm$0.7 &   24$\pm$2 \\
$[$N {\sc ii}$]$ $\lambda$6583  & 6740 & 16.6$\pm$1.7 &   56$\pm$6  \\
$[$S {\sc ii}$]$ $\lambda$6716  & 6880 &  2.1$\pm$0.4 &  6.8$\pm$1.4 \\
$[$S {\sc ii}$]$ $\lambda$6731  & 6890 &  2.9$\pm$0.6 &  9.2$\pm$1.8 \\

 & & & \\

\multicolumn{4}{c}{HD 306414} \\
                                                                    
H$_\alpha$                      & 6566 & 1.9$\pm$0.1 & 530$\pm$30 \\
                                                                 
 & & & \\

\multicolumn{4}{c}{HD 100199} \\
                                                                  
H$_\beta$                       & 4862 & 0.38$\pm$0.06 & 970$\pm$150 \\
He I $\lambda$5015$^{a}$        & 5012 & 0.09$\pm$0.02 & 220$\pm$60 \\
				& 5020 & 0.13$\pm$0.03 & 310$\pm$80 \\
He I $\lambda$5875$^{a}$        & 5871 & 0.42$\pm$0.04 & 670$\pm$70 \\
				& 5881 & 0.53$\pm$0.05 & 840$\pm$80 \\
H$_\alpha$                      & 6563 &  7.9$\pm$0.2 & 9300$\pm$300 \\
He I $\lambda$6678$^{a}$        & 6672 & 0.17$\pm$0.03 & 190$\pm$40 \\
				& 6684 & 0.24$\pm$0.05 & 270$\pm$50 \\
He I $\lambda$7065$^{a}$        & 7060 & 0.50$\pm$0.05 & 490$\pm$70 \\
				& 7070 & 0.53$\pm$0.05 & 520$\pm$50 \\
                                                                   
 & & & \\

\multicolumn{4}{c}{2RXP J130159.6$-$635806} \\

H$_\alpha$                      & 6562 & 21.7$\pm$1.1 & 2.37$\pm$0.12 \\

 & & & \\

\multicolumn{4}{c}{4U 1344$-$60} \\

H$_\alpha$ + $[$N {\sc ii}$]$$^*$ & 6650 & 259$\pm$13 & 3040$\pm$150 \\
$[$S {\sc ii}$]$ $\lambda$6731    & 6821 &  3$\pm$1 & 32$\pm$10 \\

 & & & \\

\multicolumn{4}{c}{HD 146803} \\
                                                                   
H$_\alpha$                      & 6563 & 8.7$\pm$0.4 & 1900$\pm$100 \\
                                                             
 & & & \\

\multicolumn{4}{c}{IGR J16482$-$3036} \\

H$_\gamma$ (broad comp.)        & 4493 &  50$\pm$10 & 170$\pm$30 \\
H$_\beta$ (narrow comp.)        & 5011 &  27$\pm$4  &   91$\pm$14 \\
H$_\beta$ (broad comp.)         & 5042 &  98$\pm$10 &  340$\pm$30 \\
$[$O {\sc iii}$]$ $\lambda$4958 & 5114 &  7.8$\pm$0.8 &   27$\pm$3 \\
$[$O {\sc iii}$]$ $\lambda$5007 & 5165 & 18.7$\pm$1.3 &   65$\pm$5 \\
$[$O {\sc i}$]$ $\lambda$6300   & 6501 &  6.7$\pm$0.7 &   23$\pm$2 \\
$[$O {\sc i}$]$ $\lambda$6363   & 6562 &  1.7$\pm$0.3 &  5.8$\pm$1.2 \\
H$_\alpha$ + $[$N {\sc ii}$]$$^*$ & 6769 & 366$\pm$18 & 1300$\pm$70 \\
$[$S {\sc ii}$]$ $\lambda$6716  & 6930 &  3.5$\pm$0.5 & 12.9$\pm$1.9 \\
$[$S {\sc ii}$]$ $\lambda$6731  & 6944 &  3.7$\pm$0.6 & 13.4$\pm$2.0 \\

\noalign{\smallskip}
\hline
\noalign{\smallskip}
\multicolumn{4}{l}{$^{a}$: double-peaked line.} \\
\multicolumn{4}{l}{$^*$: these lines are heavily blended. The wavelength} \\
\multicolumn{4}{l}{of the emission peak is reported.} \\
\noalign{\smallskip}
\hline
\noalign{\smallskip}
\end{tabular}
\end{center}
\end{table}

\subsection{IGR J10404$-$4625 (=LEDA 93974)}

The spectrum of the galaxy LEDA 93974 shows a 
number of narrow emission features that can readily be identified with 
redshifted optical nebular lines. These lines include [O~{\sc iii}] 
$\lambda\lambda$4959, 5007, [O~{\sc i}] $\lambda$6300, H$_\alpha$, [N~{\sc 
ii}] $\lambda\lambda$6548, 6583, and [S~{\sc ii}] $\lambda\lambda$6716, 
6731. All identified emission lines yield a redshift $z$ = 
0.0237$\pm$0.0006, in perfect agreement with the value of Paturel et al. 
(2003). The NaD doublet and the Mg b band, both in absorption, are also 
detected at the same redshift. The H$_\beta$ emission line is not detected 
down to a 3$\sigma$ flux limit of 1.4$\times$10$^{-15}$ erg cm$^{-2}$ 
s$^{-1}$.

The presence of only narrow emission lines in the optical spectrum of LEDA 
93974 suggests activity typical of a 
Narrow-Line Region (NLR) surrounding an obscured AGN (see, e.g., the 
classification of Veilleux \& Osterbrock 1987); this is also supported by 
the SUMSS radio detection and by the nondetection in the soft X--ray band. 
We further confirm this by using the diagnostic line ratios [N~{\sc 
ii}]/H$_\alpha$, [S~{\sc ii}]/H$_\alpha$, and [O~{\sc iii}]/H$_\beta$, 
together with the detection of substantial [O~{\sc i}] $\lambda$6300 
emission: the values of these quantities place this source in the regime 
of Seyfert 2 AGNs (Ho et al. 1993, 1997).

The lower limit on the H$_\alpha$/H$_\beta$ line ratio can be used to
give an estimate of the lower limit to the extinction within the NLR
itself. This line ratio, once corrected for Galactic absorption
assuming (from Schlegel et al.  1998) $E(B-V)$ = 0.16 along the
LEDA 93974 line of sight, is $>$16. Considering an intrinsic Balmer
decrement H$_\alpha$/H$_\beta$ = 2.86 (Osterbrock 1989) and the
extinction law by Cardelli et al. (1989) for the reddening within the
NLR of LEDA 93974, the observed lower limit for the
H$_\alpha$/H$_\beta$ flux ratio implies an internal $E(B-V)$ $>$ 1.8
and an extinction $A_V$ $>$ 5.6 (in the galaxy rest
frame). Substantial absorption within the NLR is also suggested by the
presence of a prominent interstellar NaD absorption doublet (with
EW$_{\rm NaD}$ = 3.1$\pm$0.3 \AA) at the redshift of the galaxy.

The lower limit on $A_V$ implies, using the empirical formula of
Predehl \& Schmitt (1995), a lower limit to the hydrogen column
density within the LEDA 93974 NLR of $N_{\rm H} >$ 1$\times$10$^{22}$
cm$^{-2}$.  Given that this is just a lower limit, and considering the
uncertainties involved in the above treatment, we chose not to apply
this correction to the following.

Using the cosmology described above we find that the luminosity distance 
to LEDA 93974 is $d_L$ = 111 Mpc, and that its X--ray luminosity is
1.1$\times$10$^{44}$ erg s$^{-1}$ in the 20--100 keV band.  This value
places the source among the most luminous Type 2 Seyfert galaxies
detected so far (cf. Malizia et al. 1999). The measured value for the
X--ray luminosity of LEDA 93974 is thus comparable to that of
``classical'' AGNs.

The optical Gunn $g$- and $r$-band magnitudes of this galaxy, reported by 
J\o rgensen et al. (1995) as $g$ = 13.55 and $r$ = 13.60, together 
with the conversion equations from these authors, allowed us to estimate 
the Johnson-Cousins $BVR$ magnitudes of LEDA 93974 as $B$ = 14.11, $V$ = 
13.55 and $R$ = 13.23. In particular, the $B$-band magnitude implies 
that the absolute optical $B$ magnitude of this galaxy is M$_B$ = $-$21.1.
This is, strictly speaking, a lower limit to the $B$-band luminosity 
of LEDA 93974, as no internal reddening was considered.

Next, assuming for IGR J10404$-$4625 an X--ray spectrum typical of a
Seyfert 2 galaxy, with a power law of photon index $\Gamma$ = 1.9 (see
Bassani et al. 1999 and references therein) we can extrapolate
the information obtained with {\it INTEGRAL} to determine the
unabsorbed 2--10 keV flux of the source to be 6.1$\times$10$^{-11}$
erg cm$^{-2}$ s$^{-1}$. The comparison between the reddening-corrected
[O~{\sc iii}] $\lambda$5007 emission flux and the 2--10 keV X--ray
flux estimated above implies an X--ray/[O~{\sc iii}]$_{\rm 5007}$
ratio of $\sim$2000, indicating that the source is well within the
Compton-thin regime for Seyfert 2 galaxies (see Bassani et al. 1999).
We however caution the reader that this is an upper limit to the
ratio, as no correction due to the absorption intrinsic to the NLR of
the galaxy was applied to the [O~{\sc iii}] $\lambda$5007 line flux.

The strength of the optical emission lines of LEDA 93974, after accounting 
for Galactic reddening, can be used to estimate the star 
formation rate (SFR) and metallicity. From the unabsorbed H$_\alpha$ flux 
in Table 2, and following Kennicutt (1998), we determine a SFR of 
0.28$\pm$0.02 $M_\odot$ yr$^{-1}$ from the reddening-corrected H$_\alpha$ 
luminosity of (3.6$\pm$0.3)$\times$10$^{40}$ erg s$^{-1}$.
We again stress that this is a strict lower limit to the SFR as the effect 
of absorption intrinsic to LEDA 93974 was not accounted for.

\subsection{HD 306414}

The spectrum of HD 306414 (see Fig. 2) is characterized by
H$_\alpha$ emission. Its presence and its EW are often found in early B
supergiants (e.g., Leitherer 1988). Besides this, and following the
spectral type classification criteria of Jaschek \& Jaschek (1987),
the narrowness of the other detected Balmer lines in absorption, the
absence of He {\sc ii} lines, and the presence of He {\sc i} and light
metal (such as C {\sc iii}) absorption lines suggest that the
star is an early B-type supergiant, in agreement with the B1\,Ia
classification of Vijapurkar \& Drilling (1993).

Although this is not sufficient to firmly identify HD 306414 as a HMXB 
optical counterpart of IGR J11215$-$5952, it is intriguing that this hard 
X--ray transient source contains in its error circle such an unusual 
emission-line object. This would be similar to the cases of other 
supergiant fast X--ray transient (SFXT) HMXBs described by Sguera et al. 
(2005) and which {\it INTEGRAL} is discovering during its Galactic Plane 
scans.

Assuming that HD 306414 is of spectral type B1\,Ia implies M$_V$ =
$-$6.4 and $(B-V)_0$ = $-$0.19 (Lang 1992). Using the observed $B$ =
10.57 and $V$ = 9.98 for this star (Klare \& Neckel 1977), gives a
color excess of $E(B-V)$ = 0.78 and a distance of $d \sim$ 6.2 kpc.
This distance estimate is in general agreement with that given by
Negueruela (2005), whereas the color excess is slightly less than,
but broadly consistent with, the value of $E(B-V)$ = 0.83 obtained
from the infrared maps of Schlegel et al. (1998), thereby supporting
the large distance.

This distance is compatible with HD 306414 being located in the far 
end of the Carina arm (see, e.g., Leitch \& Vasisht 1998) and implies,
if this star is indeed the optical counterpart of IGR J11215$-$5952,
a 20--60 keV peak luminosity of $\sim$4$\times$10$^{36}$ erg 
s$^{-1}$. This again compares well with the outburst peak 
luminosities of SFXTs (Sguera et al. 2005; Smith et al. 2005).

Thus, there is strong circumstantial evidence that IGR J11215$-$5952 and HD 
306414 may be the same object, and that this system is a fast transient 
HMXB. Clearly, pointed soft X--ray observations obtainable with satellites 
affording arcsecond localizations (such as {\it Chandra}, {\it XMM-Newton} 
or {\it Swift}) will allow a confirmation (or denial) of the above 
association.

\subsection{HD 100199}

Our observation confirms the emission-line nature of HD 100199 by the 
detection of H$_\alpha$, H$_\beta$ and several He {\sc i} lines in 
emission (see Fig. 2 and Table 2).  Again using the criteria of Jaschek \& 
Jaschek (1987), the width of the Balmer lines in absorption and the 
detection of He {\sc i} and light metal absorption lines are consistent 
with the most recent spectral type classification for HD 100199 of 
B0\,IIIe (Garrison et al. 1977).

From a closer inspection of the optical spectrum, we found that, whereas 
the Balmer emission lines are single-peaked, the He {\sc i} emission lines 
are double-peaked. These emission line properties are typical of 
early Be stars. Such objects frequently display double peaked emission 
lines in optically thin transitions, such as He {\sc i}, and single-peaked 
Balmer lines, which are optically thick. This effect is related to 
electron scattering of photons in the Be star disk-like circumstellar 
envelope and to the geometry of the envelope itself; see, e.g., Hanuschik 
et al. (1996) for an extensive atlas of line profiles, and references 
therein for details on line formation mechanisms.

We can estimate the distance to HD 100199 in the same way as performed 
above for HD 306414. The B0\,III spectral type implies M$_V$ = $-$5.1 
and $(B-V)_0$ = $-$0.295 (Lang 1992). From the observed $B$ = 8.24 and $V$ = 
8.23 of Fernie (1983) we determine a distance of $d \sim$ 3 kpc and 
$E(B-V)$ = 0.31 along the HD 100199 line of sight. This color excess is 
substantially lower than the Galactic value of $E(B-V)$ = 3.87 (Schlegel 
et al. 1998). This supports a location for HD 100199 in the near side of 
the Carina arm (see, e.g., Leitch \& Vasisht 1998).

If HD 100199 is the actual optical counterpart of IGR J11305$-$6256, then 
this distance implies a 20--60 keV luminosity of $\sim$1$\times$10$^{35}$ 
erg s$^{-1}$. This luminosity is comparable to that of the persistent Be/X 
HMXB system X Per/4U 0352+309, which moreover has a similar optical 
companion (e.g., Haberl et al. 1998). However, the lack of further 
information on the X--ray spectrum of IGR J11305$-$6256 prevents us from a 
deeper comparison between these two X--ray sources.

As for the case of HD 306414 (see Sect. 4.2), the current evidence is
insufficient to firmly identify HD 100199 as the true optical
counterpart of IGR J11215$-$5952, and that this X--ray source is a
HMXB. We therefore regard the association between these two objects as
possible, but not firm due to the lack of more multiwavelength
information.

\subsection{2RXP J130159.6$-$635806}

The brighter of the two possible counterparts within the 2RXP
J130159.6$-$635806 {\it XMM-Newton} error circle has the spectrum of a
normal M--type star, and can thus be discarded as a candidate for the
X--ray emission.  However, the spectrum of the fainter of the two
shows a prominent narrow H$_\alpha$ line at redshift zero superimposed
on a very reddened continuum (see Fig. 2). This demonstrates that this
object is the optical counterpart of this {\it INTEGRAL} hard X--ray
source, and that it is a Galactic object, most probably an accreting
X--ray binary embedded in the Galactic Plane.  This explains the very
reddened optical spectrum of this source, which shows no signal below
$\sim$5000 \AA.

The low S/N of this spectrum does not allow us to determine the
presence of any other feature, and we are thus not able to define the
spectral type of the mass donor star, and in turn its
distance. Optical photometry can nevertheless help us in this
task. Our $VRi$ data show that this object has $V$ = 19.67$\pm$0.02,
$R$ = 17.70$\pm$0.02 and $i$ = 16.08$\pm$0.03. These values, together
with the near-infrared $JHK$ magnitudes reported by Chernyakova et
al. (2005), are consistent with those of a heavily reddened X--ray
binary.

The possibility that this object is an LMXB or a CV is, however,
unlikely.  For it to be an LMXB, assuming M$_V \sim$ 0 and $(V-R)_0
\sim$ 0 (van Paradijs \& McClintock 1995), would require A$_V \sim$ 11
and a distance $d \sim$ 560 pc, too close to explain the large
reddening observed.  An even more extreme situation applies for a CV
interpretation: given M$_V \sim$ 9 and $(V-R)_0 \sim$ 0 (Warner 1995),
we obtain a distance $d \sim$ 9 pc. Thus we discard both the CV and
LMXB interpretations for this source.

If instead an early-type star is considered as the mass donor in this
X--ray binary, we find, assuming the Galactic absorption law of
Cardelli et al. (1989) and using the dereddened color indices of
Wegner (1994) for early-type stars, that the reddening towards 2RXP
J130159.6$-$635806 is $A_V \sim$ 10.5 or 9.5 for the cases of an O5\,Ia 
or B9\,V star respectively. These two extremes imply the following
distance estimates: given M$_V$ of $-$6.6 and +0.2, respectively (Lang
1992), we find $d \sim$ 14 kpc and $d \sim$ 1 kpc for O5\,Ia and 
B9\,V stars, respectively.

Neither of these extremes is, however, likely, because they are either
too far to belong to the Galactic disk (see Fig. 1 of Leitch \&
Vasisht 1998) or (again) too close to explain the large reddening
observed. Moreover, the H$_\alpha$ EW appears too large if the
secondary star is a supergiant (see Leitherer 1988). Thus, 2RXP
J130159.6$-$635806 is likely to be a HMXB hosting a late O-/early
B-type star of {\it intermediate} luminosity class.

All of the above is fully consistent with the X--ray picture of this
source reported by Chernyakova et al. (2005), who have shown that this
system hosts a slow X--ray pulsar (with spin period of $\sim$700 s) and 
that it has the characteristics of a persistent Be/X--ray binary, similar 
to the X Per/4U 0352+309 system (e.g., Haberl et al. 1998).

The following considerations can help us to better constrain the
companion spectral type. The observed optical extinction is fully
consistent with the column density of $N_{\rm H}$ =
1.7$-$1.9$\times$10$^{22}$ cm$^{-2}$ along this line of sight (Dickey
\& Lockman 1990; Chernyakova et al. 2005): using the empirical formula
of Predehl \& Schmitt (1995), an $A_V \approx$ 10 implies that $N_{\rm
H} \approx$ 1.8$\times$10$^{22}$ cm$^{-2}$.  This suggests that 2RXP
J130159.6$-$635806 is at a large distance, and that it lies within the
Crux arm, as hypothesized by Chernyakova et al. (2005). In this case,
at a distance $d \sim$ 7 kpc, we find M$_V \sim$ $-$4.5, corresponding
to either a B1\,III or a O9\,V spectral type companion (Lang 
1992), with a slight preference for the latter in terms of dereddened 
color indices.
This distance also implies a 20--100 keV luminosity of
1.6$\times$10$^{35}$ erg s$^{-1}$, which is compatible with 2RXP
J130159.6$-$635806 being a persistent HMXB similar to X Per.

Independent support for the above reddening estimates comes from the 
application of the H$\alpha$/H$\beta$ flux ratio criterion (Osterbrock 
1989) used in Sect. 4.1. The 3$\sigma$ upper limit of 6$\times$10$^{-17}$ 
erg cm$^{-2}$ s$^{-1}$ for the H$\beta$ flux implies a line ratio $>$40; 
this converts into a lower limit of A$_V >$ 8.4.

\subsection{4U 1344-60}

Inspection of the spectrum (in Fig. 2) of the fainter of the two
candidates shows that the only outstanding feature present is a large
and broad emission line superimposed on a low S/N continuum. We
readily identify this feature as the H$_\alpha$+[N {\sc ii}] complex
at a redshift $z$ = 0.013$\pm$0.001. A lower significance emission
line is also detected and identified as [S {\sc ii}] $\lambda$6731 at
the same redshift. This allows us to identify this source as a Type 1
Seyfert galaxy (according to the classification of Veilleux \&
Osterbrock 1987). No continuum emission is detected below $\sim$5500
\AA~due to the strong Galactic reddening along this line of sight: the
Galactic color excess along this direction is indeed $E(B-V)$ = 2.90
(Schlegel et al. 1998).

The spectrum of the other putative counterpart is instead basically
featureless, with a hint of H$_\alpha$ and H$_\beta$ absorption lines
at $z$ = 0. We thus identify this object as a probable normal galactic
star, and we regard the fainter of the two objects (with $R$ =
18.73$\pm$0.03, from photometry on the ESO image) as the actual 
optical counterpart of 4U 1344$-$60. 

The identification of 4U 1344$-$60 as a Type 1 Seyfert galaxy is
also supported by {\it ASCA} results (Tashiro et al. 1998), which
indicate an X--ray spectrum typical of this class of objects
(cf. Mushotzky et al. 1993). The redshift measured above corresponds
to a luminosity distance $d_{\rm L}$ = 60.6 Mpc; this in turn implies for 
4U 1344$-$60 X--ray luminosities of 1.0$\times$10$^{43}$ erg s$^{-1}$ and
3.2$\times$10$^{43}$ erg s$^{-1}$ in the 2--6 keV and 20--100 keV
bands, respectively.  This distance also implies M$_R$ = $-$22.5, although
strictly speaking this is a lower limit, as no absorption internal to
the AGN host galaxy was considered.  These X--ray and optical
luminosity estimates place 4U 1344$-$60 amongst the average of typical
Seyfert 1 AGNs (e.g., Malizia et al. 1999).

\subsection{HD 146803}

The most relevant feature in the optical spectrum of HD 146803 (see
Fig. 2) is the H$_\alpha$ profile, with a prominent emission core and
absorption wings. The other Balmer lines, in absorption, appear to be
much wider than those of supergiant stars such as HD 306414: this
suggests that the luminosity class of this star is indeed between III
and V. Moreover, the detection of He {\sc i} absorption and the
nondetection of light metal lines, apart from Mg {\sc ii} $\lambda$4481, 
suggests that this is a fairly luminous late B-type star, rather than an 
early A-type. Thus, again using the criteria of Jaschek \& Jaschek (1987), 
we suggest that HD 146803 is more likely a late-type B star of luminosity 
class III.

Assuming a spectral type of B8\,III, we can again compute its distance 
with the same method applied in Sects. 4.2 and 4.3: such stars have M$_V$ 
= $-$1.2 and $(B-V)_0$ = $-$0.11 (Lang 1992). From the observed magnitudes 
of $B$ = 10.41 and $V$ = 10.45 (see Sect. 2) we obtain a distance of 
$d \sim$ 1.9 kpc, and a color excess of $E(B-V)$ = 0.07 along the HD 
146803 line of sight. This color excess is, as for HD 100199 (Sect. 4.3),
substantially less than that, $E(B-V)$ = 5.63, obtained from the maps
of Schlegel et al. (1998), which supports the lower distance.

This places HD 146803 in the Sagittarius arm of the Galaxy (see, e.g., 
Leitch \& Vasisht 1998). On the assumption that this star is the optical 
counterpart of IGR J16207$-$5129, the above distance estimate implies a 
20--40 keV band luminosity of $\sim$1.3$\times$10$^{34}$ erg s$^{-1}$.

If HD 146803 and IGR J16207$-$5129 are the same source, the aforementioned 
spectral type identification and X--ray luminosity determination suggest 
that this object may be a low-luminosity, persistently-emitting HMXB  
with late-type B companion, similar to 1H 0739$-$529 (Liu et al. 2000 and 
references therein).  However, the nondetection of this source in the 
0.16--4.0 keV band (Grillo et al. 1992) suggests that it may be transient. 
In summary, the association between HD 146803 and IGR J16207$-$5129 is 
still tentative and further multiwavelength investigations are needed to 
confirm the identification.

\subsection{IGR J16482$-$3036}

In the spectrum of this object the most striking 
spectral feature is a prominent and broad, redshifted H$_\alpha$+[N {\sc 
ii}] emission blend topped by a narrow H$_\alpha$ emission component. A broad 
plus narrow H$_\beta$ and broad H$_\gamma$ emissions, the former flanked 
by narrow [O~{\sc iii}] emission, are also detected. Possibly, 
He~{\sc i} $\lambda$5875 is also present in emission. All of the 
narrow features have a redshift $z$ = 0.0313$\pm$0.0006, whereas that of
the broad lines is $z$ = 0.036$\pm$0.001; this kind of velocity shift 
between broad and narrow line components is not uncommon in Type 1 AGNs 
(e.g., Sulentic et al. 2000). All of the above information implies that 
this source is a Type 1 Seyfert galaxy (see Veillet \& Osterbrock 1987).

Assuming the redshift obtained from the narrow lines and the cosmology
described above, we obtain for IGR J16482$-$3036 a luminosity distance
of 146 Mpc and X--ray luminosities of 2.1$\times$10$^{42}$ erg
s$^{-1}$ and 6.9$\times$10$^{43}$ erg s$^{-1}$ in the 0.1--2.4 keV and
20--100 keV bands, respectively.  These X--ray luminosity estimates
place IGR J16482$-$3036 on the bright side of the Seyfert 1 galaxies
distribution (Malizia et al. 1999).  Analogously, this distance
implies M$_B$ = $-$20.0, again a lower limit to the $B$-band
luminosity of IGR J16482$-$3036, as no absorption internal to the AGN
host galaxy was accounted for.

Concerning this issue, the complex structure of the H$_\alpha$ plus [N
{\sc ii}] emissions does not allow us to obtain a reliable measure of
the H$_\alpha$ narrow component flux. Thus, in principle, we are not
able to correct the optical spectrum of IGR J16482$-$3036 for any
absorption that is intrinsic to the galaxy. However, assuming the same
gaussian shape for the narrow Balmer components, we obtain an
H$_\alpha$/H$_\beta$ flux ratio equal to the line peak
ratio. Therefore, given that the latter ratio is $\sim$2.7, we can
assume that no substantial further absorption is present within the
galaxy, in agreement with its Seyfert 1 nature.

For this hypothesis of negligible internal absorption, and assuming an
H$_\alpha$ narrow-component flux of 2.6$\times$10$^{-13}$ erg
cm$^{-2}$ s$^{-1}$, we can estimate the SFR in this galaxy. Again
using Eq. (1) of Kennicutt (1998) and the luminosity distance derived
above, we find a SFR of $\sim$5 $M_\odot$ yr$^{-1}$.

The complexity of the H$_\alpha$ plus [N {\sc iii}] emission region,
together with the absence of [O {\sc ii}] within our spectral range,
does not allow us to compute the metallicity of this galaxy. With the
available spectral data only an approximate lower limit of
$\approx$7.0 for the 12 + [O/H] parameter, corresponding to a metal
abundance of $\approx$0.01 $Z_\odot$, can be obtained using the
R$_{\rm 23}$ parameter as described in Kobulnicky et al. (1999).

Next, following Kaspi et al. (2000) and Wu et al. (2004), we can
compute an estimate of the mass of the central black hole in this
active galaxy.  This can be achieved using (i) the H$_\beta$ emission
flux, corrected for a foreground Galactic color excess of $E(B-V)$ =
0.34 (Schlegel et al. 1998) and (ii) a broad-line region (BLR) gas
velocity $v_{\rm BLR} \sim$ ($\sqrt{3}$/2)$\cdot$$v_{\rm FWHM}$ $\sim$
6600 km s$^{-1}$ (where $v_{\rm FWHM}$ $\sim$ 7700 km s$^{-1}$ is the
rest-frame velocity measured from the H$_\beta$ emission FWHM). Using
the above information, from Eq. (2) of Wu et al. (2004) we find that
the BLR size is $R_{\rm BLR} \sim$ 22 light-days. Furthermore, using
Eq. (5) of Kaspi et al. (2000), the AGN black hole mass in IGR
J16482$-$3036 is $M_{\rm BH} \sim$ 1.4$\times$10$^{8}$ $M_\odot$.

\section{Conclusions}

We have presented the results of the third stage (the first one dealing 
with southern objects), this time accomplished at SAAO and at ESO, of 
our ongoing observational campaign aimed at the identification of 
newly-discovered {\it INTEGRAL} objects of unknown nature. We 
spectroscopically observed the putative optical counterparts of seven 
southern {\it INTEGRAL} sources. For two cases, optical photometry was 
also obtained. This approach, already very successful as demonstrated in 
Papers I and II, allowed us to firmly determine the nature of at least 
four of these objects.

We found that: (i) IGR J10404$-$4625 (=LEDA 93974) is a Seyfert 2 galaxy
in the Compton-thin regime at $z$ = 0.0237; (ii) 2RXP 
J130159.6$-$635806 is a non-supergiant HMXB in the Crux Arm, 
at a distance $\approx$7 kpc; (iii) 4U 1344$-$60 is a Seyfert 1 galaxy 
at $z$ = 0.013; and (iv) IGR J16482$-$3036 is a Seyfert 1 galaxy 
at $z$ = 0.0313 with a central black hole of mass 
$\sim$1.4$\times$10$^{8}$ $M_\odot$.

We also give possible identifications for three further cases, namely: (i) 
HD 306414 is the likely counterpart of IGR J11215$-$5952 and thus probably 
a SFXT HMXB located at a distance $d \sim$ 6.2 kpc; (ii) HD 100199 is 
possibly associated with the hard X--ray source IGR J11305$-$6256, which, 
if correct, would then resemble the persistent HMXB X Per, but at a 
distance $\sim$3 kpc; (iii) HD 146803 is tentatively associated with IGR 
J16207$-$5129; which would identify it as a HMXB, possibly similar to 
1H 0739$-$529, and located at a distance of $\sim$1.9 kpc. However, more 
extensive multiwavelength studies of the error boxes of these sources, 
especially through pointed soft X--ray observations (obtainable with 
satellites such as {\it Chandra}, {\it XMM-Newton} or {\it Swift}), will 
allow a conclusive test of these three possible associations.

Moreover, all of the above once again demonstrates, as already
remarked in Papers I and II, that {\it INTEGRAL} is making a
fundamental contribution in the detection and study of a
substantial fraction of persistent and transient HMXBs along the
Galactic Plane, and of background AGNs located beyond the Zone of
Avoidance of the Galactic Plane.

\begin{acknowledgements}

We thank I. Saviane for assistance with the EFOSC2 spectroscopic
calibration frames.  This research has made use of data retrieved from
the ESO/ST-ECF Science Archive; it has also made use of the NASA
Astrophysics Data System Abstract Service, of the NASA/IPAC
Extragalactic Database (NED), and of the NASA/IPAC Infrared Science
Archive, which are operated by the Jet Propulsion Laboratory,
California Institute of Technology, under contract with the National
Aeronautics and Space Administration. We also acknowledge the SIMBAD
database operated at CDS, Strasbourg, France, and the HyperLeda
catalogue operated at the Observatoire de Lyon, France.
A referee is thanked for some useful comments which helped us to 
improve the quality of this paper.

\end{acknowledgements}

\noindent
{\bf Note added in proof.}
Recent (24 December 2005) publicly-available observations of the field of
IGR J11305$-$6256, performed with the XRT instrument onboard the
{\it Swift} satellite, showed that the only X--ray source detected within
the ISGRI error box is positionally fully consistent with star HD 100199.
The accurate X--ray localization (J2000; RA = 11 31 06.5, Dec =
$-$62$^{\circ}$ 56$'$ 46$\farcs$6, error radius: 6$''$) afforded with XRT
lies 3$\farcs$5 from HD 100199 and thus proves beyond any resonable doubt
that, indeed, this star is the optical counterpart of IGR J11305$-$6256.

\end{document}